\documentclass[prb,superscriptaddress,amsmath,amssymb,twocolumn,floatfix]{revtex4}
\usepackage{natbib}
\usepackage{graphicx}         
\usepackage{dcolumn}          
\usepackage{bm}               
\usepackage{upgreek}
\usepackage{booktabs} 				
\usepackage{tabularx}
\usepackage{array}
\usepackage[colorlinks=true,urlcolor=blue,breaklinks=true]{hyperref}

\begin{document}

\title{Flux periodic oscillations and phase-coherent transport in GeTe nanowire-based devices}

\author{Jinzhong Zhang}
\affiliation{Peter Gr\"unberg Institut (PGI-9), Forschungszentrum J\"ulich, 52425 J\"ulich, Germany}
\affiliation{JARA-Fundamentals of Future Information Technology, J\"ulich-Aachen Research Alliance, Forschungszentrum J\"ulich and RWTH Aachen University, Germany}
\affiliation{Technical Center for Multifunctional Magneto-Optical Spectroscopy (Shanghai), Engineering Research Center of Nanophotonics \& Advanced Instrument (Ministry of Education), Department of Materials, School of Physics and Electronics Science, East China Normal University, Shanghai 200241, China.}

\author{Pok-Lam Tse}
\affiliation{Department of Physics and Astronomy and Department of Electrophysics, University of Southern California, CA 90089, Los Angeles, USA}

\author{Abdur-Rehman Jalil}
\affiliation{Peter Gr\"unberg Institut (PGI-9), Forschungszentrum J\"ulich, 52425 J\"ulich, Germany}
\affiliation{JARA-Fundamentals of Future Information Technology, J\"ulich-Aachen Research Alliance, Forschungszentrum J\"ulich and RWTH Aachen University, Germany}

\author{Jonas K\"{o}lzer}
\affiliation{Peter Gr\"unberg Institut (PGI-9), Forschungszentrum J\"ulich, 52425 J\"ulich, Germany}
\affiliation{JARA-Fundamentals of Future Information Technology, J\"ulich-Aachen Research Alliance, Forschungszentrum J\"ulich and RWTH Aachen University, Germany}

\author{Daniel Rosenbach}
\affiliation{Peter Gr\"unberg Institut (PGI-9), Forschungszentrum J\"ulich, 52425 J\"ulich, Germany}
\affiliation{JARA-Fundamentals of Future Information Technology, J\"ulich-Aachen Research Alliance, Forschungszentrum J\"ulich and RWTH Aachen University, Germany}

\author{Martina Luysberg}
\affiliation{Ernst Ruska Center, Forschungszentrum J\"ulich, 52425 J\"ulich, Germany}

\author{Gregory Panaitov}
\affiliation{Institute of Complex Systems (ICS-8) Forschungszentrum J\"ulich, 52425 J\"ulich, Germany}

\author{Hans L\"uth}
\affiliation{Peter Gr\"unberg Institut (PGI-9), Forschungszentrum J\"ulich, 52425 J\"ulich, Germany}
\affiliation{JARA-Fundamentals of Future Information Technology, J\"ulich-Aachen Research Alliance, Forschungszentrum J\"ulich and RWTH Aachen University, Germany}

\author{Zhigao Hu}
\affiliation{Technical Center for Multifunctional Magneto-Optical Spectroscopy (Shanghai), Engineering Research Center of Nanophotonics \& Advanced Instrument (Ministry of Education), Department of Materials, School of Physics and Electronics Science, East China Normal University, Shanghai 200241, China.}

\author{Detlev Gr\"{u}tzmacher}
\affiliation{Peter Gr\"unberg Institut (PGI-9), Forschungszentrum J\"ulich, 52425 J\"ulich, Germany}
\affiliation{JARA-Fundamentals of Future Information Technology, J\"ulich-Aachen Research Alliance, Forschungszentrum J\"ulich and RWTH Aachen University, Germany}

\author{Jia Grace Lu}
\affiliation{Department of Physics and Astronomy and Department of Electrophysics, University of Southern California, CA 90089, Los Angeles, USA}

\author{Thomas Sch\"apers}
\email{th.schaepers@fz-juelich.de (Th. Sch\"apers)}
\affiliation{Peter Gr\"unberg Institut (PGI-9), Forschungszentrum J\"ulich, 52425 J\"ulich, Germany}
\affiliation{JARA-Fundamentals of Future Information Technology, J\"ulich-Aachen Research Alliance, Forschungszentrum J\"ulich and RWTH Aachen University, Germany}

\hyphenation{na-no-struc-tures na-no-scale spin-tro-nics na-no-elec-tron-ics}
\date{\today}

\begin{abstract}
Despite the fact that GeTe is known to be a very interesting material for applications in thermoelectrics and for phase-change memories, the knowledge on its low-temperature transport  properties is only limited. Here, we report on phase-coherent phenomena in the magnetotransport of GeTe nanowires. From universal conductance fluctuations, a phase-coherence length of about 200\,nm at 0.5\,K is determined for the hole carriers. The distinct phase-coherence is confirmed by the observation of Aharonov--Bohm type oscillations for magnetic fields applied along the nanowire axis. We interpret the occurrence of these magnetic flux-periodic oscillations by the formation of a tubular hole accumulation layer on the nanowire surface. In addition, for Nb/GeTe-nanowire/Nb Josephson junctions, we obtained a proximity-induced critical current of about 0.2\,$\mu$A at 0.4\,K. By applying a magnetic field perpendicular to the nanowire axis, the critical current decreases monotonously with increasing magnetic field, which indicates that the structure is in the small-junction-limit. Whereas, by applying a parallel magnetic field the critical current oscillates with a period of the magnetic flux quantum indicating once again the presence of a tubular hole channel. 
\end{abstract}
\maketitle

\section{Introduction}

In the past decades, GeTe-based alloys are attracting an increasing interest for their applications in optical data storage and phase change memories due to a reversible rapid transformation between amorphous and crystalline phase\cite{MWuttigNM2007,DLencerNM2008} and a large contrast of optical constants and electrical conductivity for the two phases.\cite{SGuoAPL2015} In addition, GeTe-rich alloys can be used in intermediate temperature thermoelectric applications for converting waste heat to electrical energy because of their high thermoelectric performance, high thermal and mechanical stability.\cite{MHongAM2019,SPerumalJMCC2016} Moreover, the GeTe ferroelectric semiconductor exhibits a giant bulk Rashba effect,\cite{GSPawleyPRL1966,LPonetPRB2018,MLiebmannAM2016,JKrempaskyPRR2020} which is very interesting for applications in spintronic devices. \cite{DDSanteAM2012,MSBahramyAM2017} 

It is well known that GeTe has different phases upon increasing the temperature or pressure.\cite{DJSinghJAP2013,MHongAM2019,APawbakePRL2019} At room temperature, GeTe is an indirect band gap semiconductor with a trigonal ferroelectric phase (space group $R3m$, No. 160). The experimentally determined band gap energy is $E_\mathrm{g}\sim$\,0.61\,eV,\cite{JWParkAPL2008} which is close to recent theoretical values.\cite{DDSanteAM2012,DJSinghJAP2013} The band gap energy tends to decrease with decreasing temperature.\cite{MHongAM2019} Pristine GeTe is a $p$-type semiconductor with hole concentration up to 10$^{21}$\,cm$^{-3}$ due to the low formation energy of Ge vacancies.\cite{MHongAM2019} At temperatures below 0.3\,K GeTe is observed to be The superconducting.\cite{RAHeinPRL1964,JLSmithJLTP1977,VNarayanpss2016} Recently, Narayan \emph{et al.}\cite{VNarayanPRB2019} have investigated the non-equilibrium superconductivity in GeTe Hall bar samples with a semiconductor-superconductor transition temperature of about 0.14\,K and a critical magnetic field of around 70\,mT. At the Curie temperature of about 700\,K, GeTe undergoes a phase transition from the ferroelectric rhombohedral ($R3m$) structure to the paraelectric cubic ($Fm\bar{3}m$) NaCl-type one. The cubic GeTe has a direct band gap with a smaller value of about 0.2\,eV.\cite{DJSinghJAP2013,RTsuPR1968} In addition, there is an amorphous (high-resistance phase) to crystalline (low-resistance phase) transition at around 200\,$^\circ$C for GeTe films upon heating.\cite{KSAndrikopoulosJPCM2006,SGuoAPL2017} For device applications investigations on GeTe nanowires were conducted.\cite{SMeisterNL2006,SMeisterNL2008,MLongoJCG2011} These nanowires are in particular interesting as building blocks in phase-change memories.\cite{SHLeeNN2007} 

The mechanism of low-temperature quantum transport phenomena in one-dimensional GeTe nanowires have not been addressed so far, even though the phase change behavior and thermoelectric properties of GeTe bulk/films have been studied systematically. Therefore, we have  grown GeTe nanowires and measured the transport properties of GeTe nanowires contacted with normal metals as well as GeTe nanowire-based Josephson junctions. For the normal contacted nanowires we investigated phase-coherent transport phenomena such as universal conductance fluctuations, weak antilocalization, as well as Aharonov--Bohm-type resistance oscillations. For the GeTe nanowires equipped with Nb contacts the proximity-induced superconductivity as well as critical temperature and magnetic field of Nb/GeTe-nanowire/Nb Josephson junctions are studied. 

\begin{figure*}
\centering
  \includegraphics[width=0.95 \textwidth]{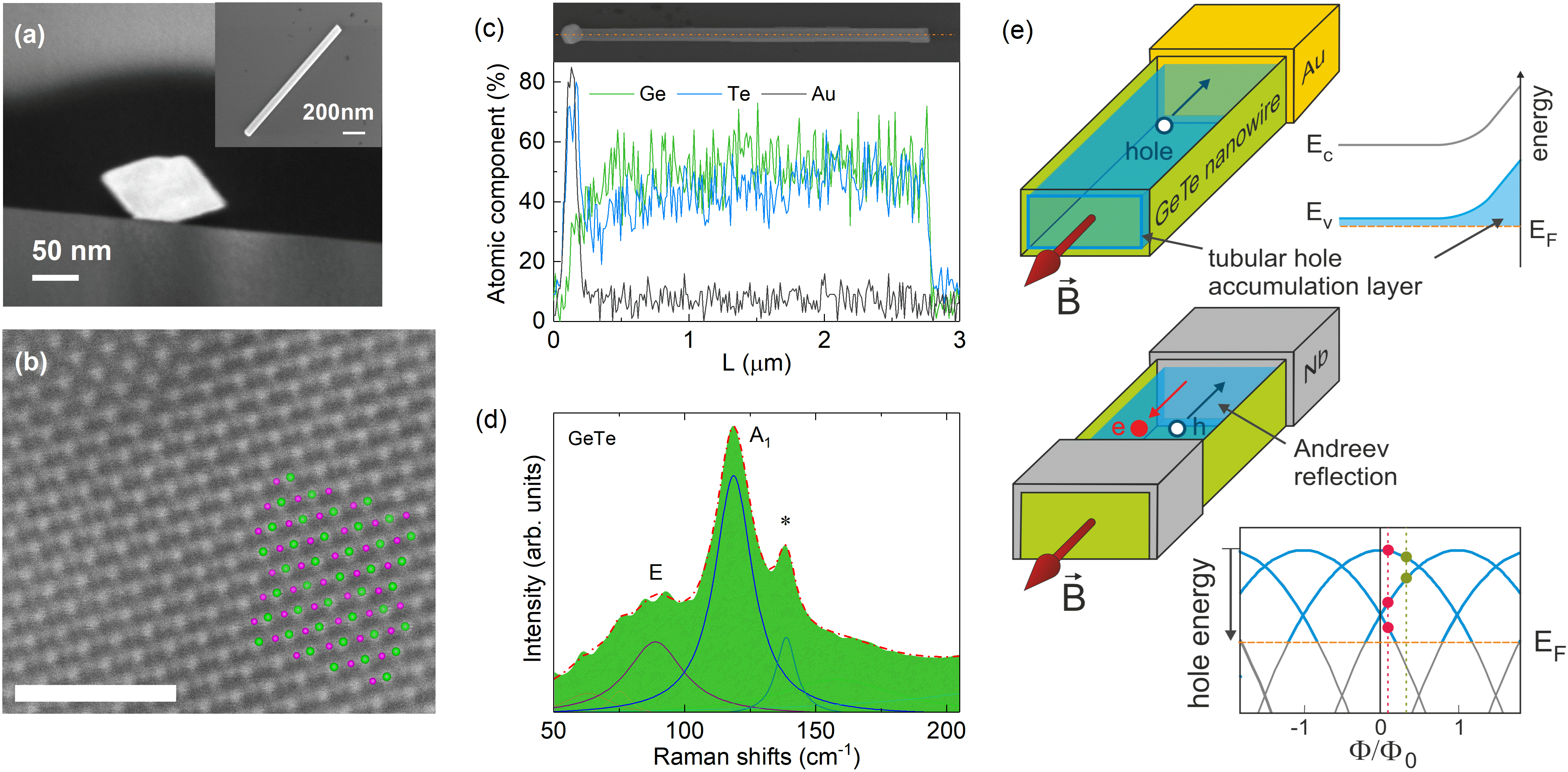}
  \caption{a) Scanning transmission electron micrograph of GeTe NW in cross section.  b) HAADF image of GeTe NW cross section obtained via aberration corrected STEM with evident trigonal crystal structure where light green atoms represent Te and magenta colored atoms indicate the position of Ge atoms.  c) A SEM image of a GeTe nanowire and the corresponding atomic composition along the nanowire from Energy dispersive X-ray spectroscopy. d) Experimental and fitted Raman spectra (solid lines) with mode assignments of a GeTe single nanowire. e) Top: Schematics of a GeTe nanowire with a tubular hole accumulation layer and a magnetic field along the wire axis. The scheme on the right shows the upwards bending of the bands at the surface, with $E_F$ the Fermi energy as well as $E_c$ and $E_v$ the conduction band bottom and valence band top, respectively. Bottom: Schematics of a GeTe nanowire contacted with Nb electrodes. At the interface, Andreev reflection occurs. The schematics on the right shows the energy spectrum of hole states in a tubular conductor for different angular momenta as a function of the normalized flux $\Phi / \Phi_0$ for $E_{kin}=0$. 
  Note that for hole states the energy increases downwards and the effective mass is positive. The red and green dots indicate the occupied channels at different flux values.}
  \label{Fig-SEM-GeTe}
\end{figure*}

\section{Results and Discussion}

\subsection{Crystalline structure and microstructure analysis}

GeTe nanowires (NWs) were synthesized by a Au-catalyzed vapor-liquid-solid growth in a tube furnace with GeTe powders.\cite{SMeisterNL2006,DYuJACS2006} The as-grown GeTe NWs on a Si/SiO$_2$ substrate are straight with a length of more than 5\,$\mu$m. The cross section for the GeTe nanowires has a rhombic shape with a side length of about 80\,nm (cf. Figure~\ref{Fig-SEM-GeTe}a). A high-angle annular dark- field (HAADF) image of GeTe NW cross section was obtained via aberration corrected scanning transmission electron microscopy (STEM). In the corresponding image shown in Fig.~\ref{Fig-SEM-GeTe}b one can identify a trigonal crystal structure. The positions of the Te and Ge atoms are indicated by light green and magenta dots, respectively. For device fabrication the GeTe NWs were transferred to a Si/SiO$_2$ substrate. Subsequently, superconducting (Nb) or normal-metal (Ti/Au) contact electrodes were prepared by using electron beam lithography and magnetic sputtering. The transport properties of Nb/GeTe/Nb and Au/GeTe/Au devices are studied in a four-terminal configuration. The outer pair of electrodes are employed to apply a direct current ($I_\mathrm{bias}$) or/and an alternating current ($I_\mathrm{AC}$), while the inner pair of leads serve as voltage probes. The contact resistance can be excluded using a real four-terminal geometry base on the differential resistance $R=dV/dI$\cite{OVSkryabinaAPL2017}. 

In order to investigate the material properties of the nanowires energy dispersive X-ray spectroscopy (EDX) and Raman spectroscopy were performed. The EDX results shown in Figure~\ref{Fig-SEM-GeTe}c reveal that the as-grown GeTe has an ideal atomic ratio of Ge\,:\,Te\,=\,1\,:\,1 and the element component are uniform along the NWs except the nanowire ends with a Au nanoparticle, typical of vapor-liquid-solid growth. Furthermore, Raman spectra of a GeTe single nanowire demonstrate unambiguously that GeTe NWs exhibit rhombohedral phase (\textit{R3m}) at room temperature (cf. Figure~\ref{Fig-SEM-GeTe}d). According to the group theory, the two peaks near 89 and 119\,cm$^{-1}$ are assigned to the first-order Raman-active modes E(TO) and A$_1$(LO), respectively.\cite{VBragagliaSR2016,RShaltafPRB2009} Theoretically, E and A$_1$ relates to the bending modes of the GeTe$_4$ tetrahedra and symmetric stretching vibration of the Te-Te bond, respectively.\cite{KSAndrikopoulosJPCM2006} Note that the peak labeled by the symbol ($*$) at around 140\,cm$^{-1}$ arises from surface oxidation.\cite{APawbakePRL2019} Other peaks/shoulders may originate from the breaking of inversion symmetry due to vacancies, defects, and distortion of bonds.\cite{AShaliniJAP2015} 

\subsection{Magnetotransports in normal-metal Au/GeTe-nanowire/Au devices}

\begin{figure*}
\centering
  \includegraphics[width=0.9 \textwidth]{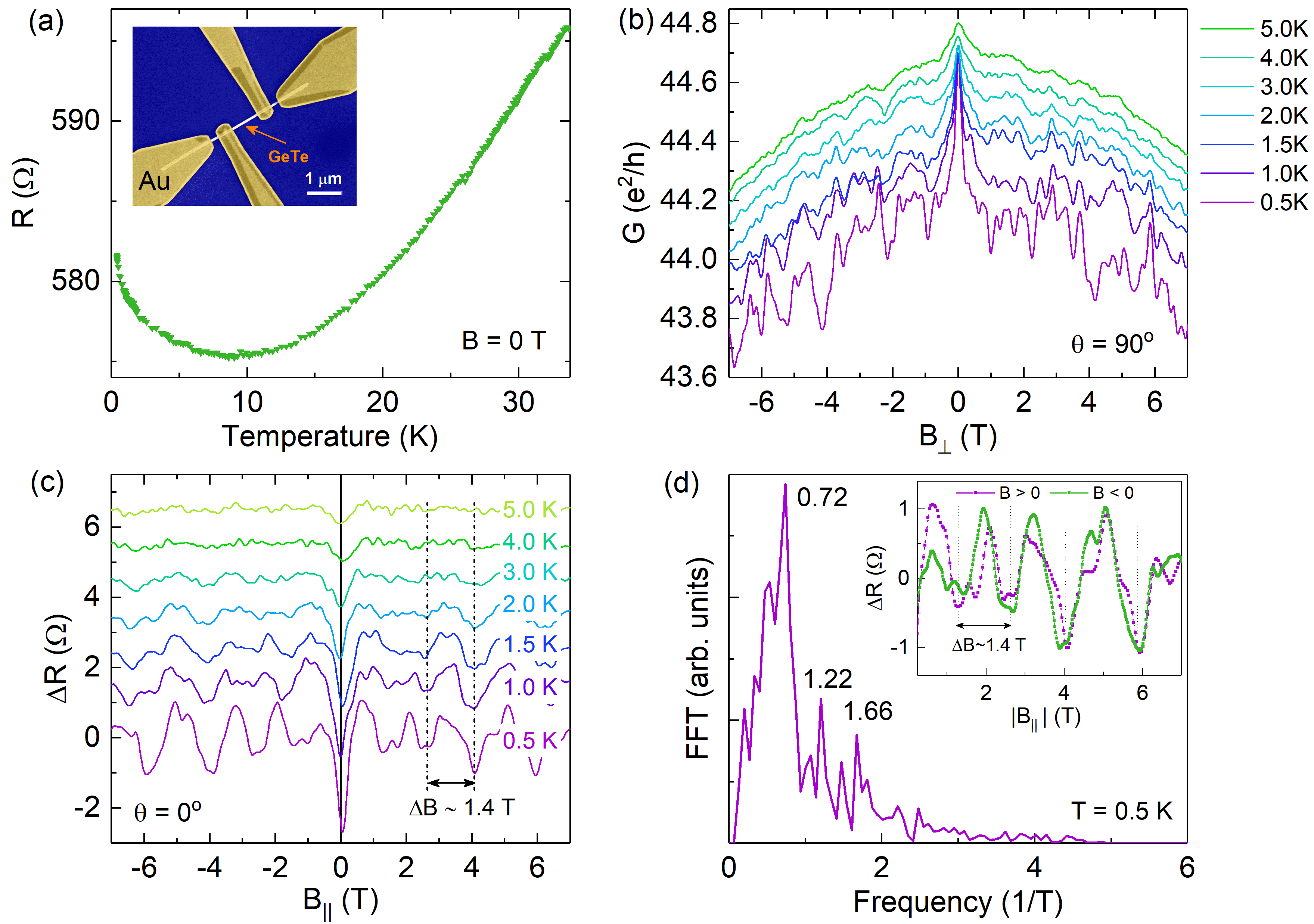}
  \caption{(Color online)
  a) Four-terminal resistance as a function of temperature at zero magnetic field of a Au/GeTe-nanowire/Au device. Inset: Scanning electron microscopy image of a GeTe NW contacted with Ti/Au leads. b) Conductance in units of $e^2/h$ as a function of $B_\perp$ at temperatures from 0.5 to 5.0 K. c) Magnetoresistance oscillations after subtracting a parabolic background as a function of B$_{||}$ at temperatures from 0.5 to 5.0 K. d) Fast Fourier transform (FFT) of the magnetoresistance oscillations in a parallel magnetic field at 0.5\,K. The inset shows the corresponding resistance oscillations for the positive and negative magnetic field range.}
  \label{Fig-R-T-GeTe-Au}
\end{figure*}

In order to gain information on the electrical properties of the GeTe nanowires we measured the magnetotransport of samples equipped with normal contacts (Ti/Au: 20\,nm/100\,nm). An electron beam micrograph of the contacted 80-nm-wide nanowire is shown in Figure~\ref{Fig-R-T-GeTe-Au}a (inset). The measurements were performed in a four-terminal configuration with the bias current supplied at the outer contacts. The width and separation of the inner electrodes is about 0.5 and 1.0\,$\mu$m, respectively. In Figure~\ref{Fig-R-T-GeTe-Au}a the resistance as a function of temperature is plotted. One finds that the resistance decreases with decreasing temperature in the temperature range from 35 to 10\,K, and then increases monotonously upon further decrease of the temperature down to 0.4\,K. The temperature dependence corresponds to a metallic behaviour where the upturn in resistance for temperatures below 10 K can be attributed to the electron-electron interaction and weak localization effects.\cite{CWeyrichJPCM2016,PALeeRMP1985,GBergmannPR1984} No transition to a superconducting behaviour is observed, probably because the lowest temperature of 0.4\,K is larger than the critical temperature.\cite{RAHeinPRL1964,VNarayanpss2016} 

Figure~\ref{Fig-R-T-GeTe-Au}b shows the conductance $G$ in units of $e^2/h$ with the magnetic field applied perpendicular to the nanowire axis (B$_{\perp}$, $\theta$\,=\,90$^\circ$), at temperatures between 0.5 and 5.0\,K. Here, $e$ is the electron charge and $h$ is the Planck constant. Similarly to GeTe films,\cite{VNarayanpss2016} a pronounced peak is observed at zero magnetic field which is attributed to the weak antilocalization effect. This feature originates from interference effects of holes in the presence of spin-orbit coupling. Indeed, for GeTe a bulk Rasbha effect in the valence band is theoretically predicted.\cite{DDSanteAM2012,MLiebmannAM2016} Furthermore, as discussed in more detail below, band bending at the surface in connection with according potential gradients may also result in spin-orbit coupling effects. The conductance traces shown in Figure~\ref{Fig-R-T-GeTe-Au}b reveal a modulation pattern which can be assigned to universal conductance fluctuations.\cite{PALeePRL1985} Their amplitude decreases with increasing the temperature mainly due to the reduction of the phase-coherence length.\cite{YCArangoCR2016} From the autocorrelation function of the magnetoconductance, given by $F(\Delta B)=\langle G(B)G(B+\Delta B)\rangle-\langle G(B)\rangle^2$, information on the phase-coherence length $l_\phi$ can be obtained.\cite{PALeePRL1985} Here, $\langle \dots \rangle$ is the average over the magnetic field. For quasi one-dimensional systems the correlation field $B_c$ defined by $F(B_c)=F(0)/2$, is inversely proportional to the phase-coherence length $l_\phi$:  $B_c \sim \sqrt{3}\Phi_0/(\pi l_\phi d)$, with the magnetic flux quantum given by $\Phi_0$\,=\,$h/e$.\cite{DKFerryCUP2009,CWJBeenakkerSSP1991} For the fluctuations measured at 0.5\,K, we extracted a correlation field of $B_c=0.14\,$T. Taking the nanowire width of $d$\,$\sim$\,80\,nm into account we obtain a phase-coherence length of about 200\,nm. 
\begin{figure*}
\centering
  \includegraphics[width=0.9 \textwidth]{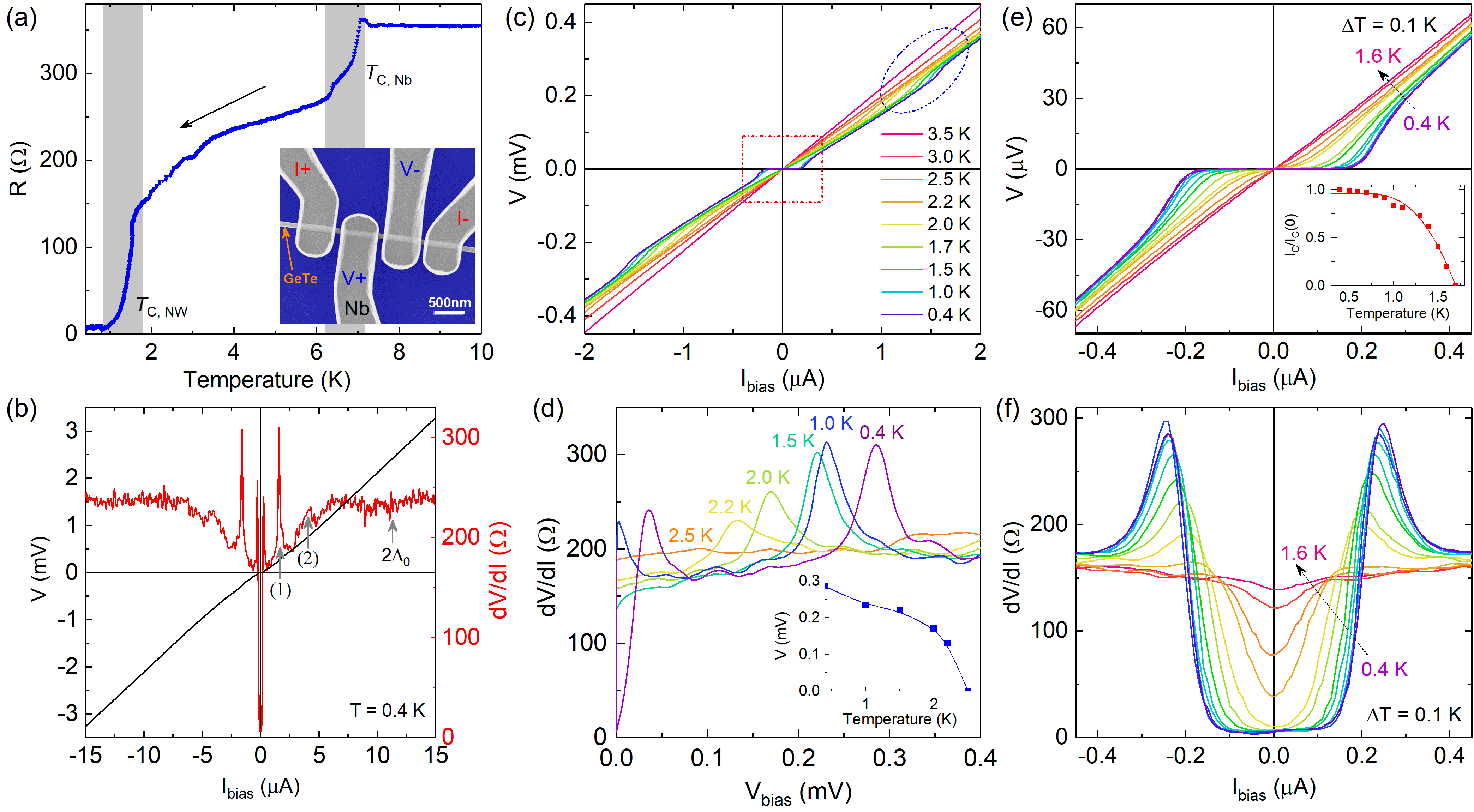}
  \caption{a) Resistance $R$ of a Nb/GeTe-nanowire/Nb junction during a cooling process. There are two transitions ($T_\mathrm{C,Nb}$ and $T_\mathrm{C,NW}$) labelled by vertical shadows. Inset: A scanning electron microscopy image of a Nb/GeTe-nanowire/Nb junction. b) \textit{I-V} characteristics and corresponding \textit{dV/dI} trace of a GeTe-based junction in a high bias current range at 0.4 K. The subgap features and the location of $2\Delta_0$ are indicated by arrows. c) \textit{I-V} curves as a function of large bias current range. The superconducting range at small bias currents is indicated by an rectangle and the step feature at around $\pm$1.5\,$\mu$A  can be found within the ellipse. d) Corresponding \textit{dV/dI} traces as a function of bias voltage at various temperatures. Inset: Position as a function of temperature of the peak assigned feature (1) indicated in b). e) \textit{I-V} characteristics and f) the corresponding \textit{dV/dI} curves in a small bias current range at temperatures between 0.4 and 1.6\,K. The inset of e) is the experimental (dots) and fit (solid line) $I_\mathrm{C}$/$I_\mathrm{C}$(0) as a function of temperature.
  }
  \label{Fig-R-T-GeTe-Nb}
\end{figure*}

Figure~\ref{Fig-R-T-GeTe-Au}c shows the resistance $\Delta R$  after subtraction of a slowly varying background as a function of a magnetic field $B_{||}$ applied along the nanowire axis. For this magnetic field orientation one finds a regular oscillation pattern with an oscillation period of about $\Delta B= 1.4$\,T (cf. Figure~\ref{Fig-R-T-GeTe-Au}d, inset). This is also confirmed by the fast Fourier spectrum of the measurement at 0.5\,K depicted in Figure~\ref{Fig-R-T-GeTe-Au}d. Here, a pronounced peak is found at a frequency of 0.72\,T$^{-1}$, corresponding to the aforementioned period of 1.4\,T. In addition, some features at 1.22 and 1.66\,T$^{-1}$ are present. 

We attribute the regular oscillations to Aharonov--Bohm type oscillations\cite{YAharonovPR1959} originating from transport in tubular hole states in a conductive surface accumulation layer penetrated by a magnetic flux $\Phi$ (cf. Fig.~\ref{Fig-SEM-GeTe}e, upper left schematics).\cite{TRichterNL2008,HPengNatMat2010,OGuelPRB2014,SChoNC2015,YCArangoCR2016} The involvement of phase-coherent tubular states is supported by the large phase-coherence length being comparable to the circumference of the nanowire. We assume that in case of our GeTe nanowire, the holes accumulated in the surface channel are supplied from surface states. In fact, such a surface accumulation layer is mainly observed for narrow band gap materials, e.g. for InAs. Similar but inverted as in the case of InAs the GeTe electronic band structure exhibits some lower valence band maxima,\cite{HMPolatoglouPNGS1982,AJainAPLMat2013} which may be responsible for the charge neutrality level of the surface states to be located within the bulk valence band. As a consequence, a tubular hole accumulation layer is formed at the surface.\cite{HLuethSpringer2015} Owing to the hole accumulation the valence band is bent upwards at the interface (cf. Fig.~\ref{Fig-SEM-GeTe} e).  
In order to describe the transport through the tubular hole channel in more detail, we assume a circular cross section. In that case the state components perpendicular to the nanowire axis can be described by coherent angular momentum states, which can be ordered according to their angular momentum quantum number $l$ of their angular momentum $L_z$. In the presence of a magnetic flux through the wire cross section the energy eigenstates are given by\cite{TRichterNL2008,OGuelPRB2014}
$$
E=E_{kin} + \frac{\hbar^2}{2m^*r_0^2} \left( l-\frac{\Phi}{\Phi_0}\right)^2 \; ,
$$
with $E_{kin}$ the kinetic energy along the wire, $m^*$ the effective hole mass, and $r_0$ the radius of the tubular channel. The energy is periodic in $\Phi_0=h/e$. A schematics of the energy spectrum is given in Figure~\ref{Fig-SEM-GeTe}e, bottom. In a ballistic picture the conductance is determined by the number of occupied hole channels above the Fermi level $E_F$, e.g. three channels for a small flux (red dots in the schematics) and two channels for a slightly larger flux (green dots).\cite{TRichterNL2008} Thus, when the magnetic field is increased, the number of occupied hole channels changes periodically with the period of $\Phi_0$. In the non-ballistic diffusive case, a similar flux-periodic modulation is expected as well.\cite{TORosdahl2014}   

From the measured period of $\Delta B= 1.4$\,T of our nanowire we can deduce the expected area $S$ encircled by the surface channel from $\Phi_0=S\times \Delta B$, which results in a value of about $S=3000\,\mathrm{nm}^2$, corresponding to a diameter of approximately 60\,nm for a circular cross section. For our nanowire with a width of $80\,$nm, we estimate an area of about $5000\,\mathrm{nm}^2$ assuming a circular cross section. Obviously, the expected cross section is smaller than the actual geometrical cross sectional area of our GeTe nanowire. The discrepancy may be attributed to the fact that the nanowire surface is oxidized so that the hole accumulation layer is slightly pushed inside. Furthermore, the radial component wave function usually has an extension of the few nanometers, which also leads to a reduction of the effective cross sectional area. As mentioned above, some weaker maxima at higher frequencies, i.e. at 1.22 and 1.66\,$\mathrm{T}^{-1}$, are found in the Fourier spectrum. These features may be related to higher harmonics or to Altshuler--Aronov--Spivak oscillations comprising a $h/2e$ periodicity.\cite{BLAltshulerJETP1981}

\subsection{Temperature dependent superconductivity in Nb/GeTe-nanowire/Nb junctions}

We now turn to measurements of GeTe nanowires equipped with superconducting contacts. The inset of Figure~\ref{Fig-R-T-GeTe-Nb}a shows a GeTe NW-based junction device with Nb contacts. The nanowire has a width of 80\,nm, while the width and separation of Nb leads are about 500 and 90\,nm, respectively. The outer Nb leads are employed to apply bias current, while the inner pair of contacts serve as voltage probes.\cite{HCourtoisPRB1995} Figure~\ref{Fig-R-T-GeTe-Nb}a shows the resistance $R$ at zero magnetic field as a function of temperature. The measurement shown in Figure~\ref{Fig-R-T-GeTe-Nb}a indicates that the Nb electrodes have a transition temperature of about $T_\mathrm{c,Nb}$\,=\,7.0\,K. From the relation of the electron-phonon coupling strength $2 \Delta_0 / k_\mathrm{B} T_\mathrm{c,Nb}$\,$\approx$\,3.9, with $k_\mathrm{B}$ the Boltzmann constant and critical temperature $T_\mathrm{c,Nb}$, we estimate the Nb energy gap $\Delta_0$ to be about 1.2\,meV at zero temperature.\cite{JPCarbotteRMP1990,HYGuenelJAP2012} Moreover, there is another drop in the temperature region between 1.0 and 2.0\,K, which is assigned to the superconducting transition temperature $T_\mathrm{c,NW}$ of the Nb/GeTe-NW/Nb junction. The two-step feature has been observed before in normal metal,\cite{LAngersPRB2008} in semiconductor-based Josephson junctions,\cite{BKKimAN2017} as well as in topological insulator based junctions.\cite{JZZhangAnnalPhys2020} The normal state four-terminal resistance ($R_\mathrm{N}$) of the present Nb/GeTe-NW/Nb junction is about 355\,$\Omega$ at temperatures above $T_\mathrm{c,Nb}$.

Figure~\ref{Fig-R-T-GeTe-Nb}b shows the current-voltage (\textit{I-V}) characteristics and corresponding differential resistance (\textit{dV/dI}) curve measured at 0.4\,K. At low bias currents a superconducting state is observed with a critical current of 200\,nA. At larger bias current at around $\pm$1.6\,$\mu$A corresponding to a bias voltage of about $\pm$0.3\,mV a sharp peak (1) is observed in the differential resistance. Moreover, a broader peak (2) at $\pm$4\,$\mu$A corresponding to a bias voltage of $\pm$0.75\,mV is found. These features can be attributed to multiple Andreev reflections.\cite{OctavioPRB1983} However, no regular sequence of structures according to the theory is observed. At a bias voltage corresponding to $2\Delta_0=2.4$\,meV no pronounced feature is found.  

As a further step, the temperature dependence of the \textit{I-V} characteristics is investigated (cf. Figure~\ref{Fig-R-T-GeTe-Nb}c). Obviously, the superconducting range at small bias currents (indicated by a rectangle) shrinks and finally vanishes upon increasing the temperature from 0.4 to 3.5\,K. The same is true for the step feature at around $\pm$1.5\,$\mu$A. Regarding the latter, the corresponding \textit{dV/dI} traces as a function of bias voltage are shown in Figure~\ref{Fig-R-T-GeTe-Nb}d for temperatures in the range from 0.4 to 2.5\,K. One finds that the feature corresponding to peak (1) in Fig.~\ref{Fig-R-T-GeTe-Nb}b shifts towards smaller bias voltages and eventually vanishes at a temperature of 2.5\,K (cf. Figure~\ref{Fig-R-T-GeTe-Nb}d, inset). The strong temperature dependence confirms that this feature is directly related to the superconducting state.  

Figure~\ref{Fig-R-T-GeTe-Nb}e shows the \textit{I-V} characteristics of the GeTe-NW based Josephson junction in a small bias current range at temperatures ranging from 0.4 to 1.6\,K. At 0.4\,K the critical current $I_\mathrm{c,NW}$ is about 200\,nA. As the temperature increases, the superconducting plateau diminishes and disappears at about 1.6\,K. The according corresponding differential resistance traces are shown in Figure~\ref{Fig-R-T-GeTe-Nb}f. The dependence of normalized critical current $I_\mathrm{c}/I_\mathrm{c}$(0) as a function of temperature are depicted in the inset of Figure~\ref{Fig-R-T-GeTe-Nb}e. The temperature dependence of the supercurrent can be assigned to a metallic diffusive junction.\cite{PDubosPRB2001} However, since no reliable data on the diffusion constant and Thouless energy can be obtained, we have refrained from a detailed analysis.    

\subsection{Magnetic field dependent superconductivity in Nb/GeTe-NW/Nb junctions}

\begin{figure*}
\centering
  \includegraphics[width=0.9\textwidth]{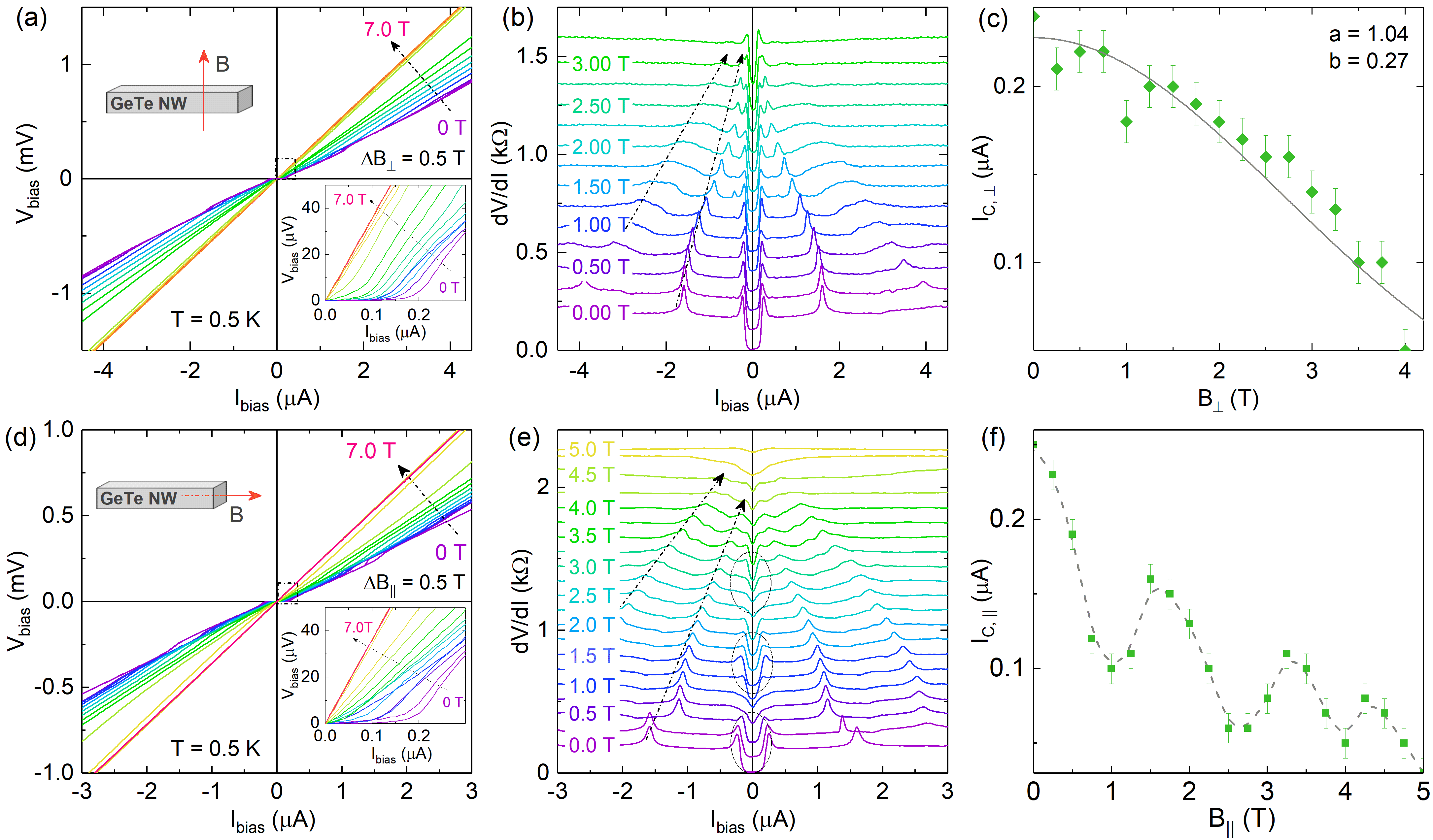}
  \caption{a) \textit{I-V} characteristics and b) the corresponding \textit{dV/dI} of a Nb/GeTe-NW/Nb junction as a function of bias current at 0.5\,K by applying various perpendicular magnetic fields $B_\perp$. The inset in a) shows a detail at small bias currents. c) Experimental (dots) and fitted (solid line) critical current I$_{\mathrm{c},\perp}$ as a function of magnetic field (B$_\perp$) based on the \textit{dV/dI} at a small bias current region (the inset). d) \textit{I-V} characteristics and e) the corresponding \textit{dV/dI} as a function of bias current at 0.5\,K by applying various magnetic field parallel to the GeTe NW axis (B$_{//}$). In e) the ranges where a supercurrent is observed are indicated by the ellipses. f) Critical current I$_{\mathrm{c},||}$ as a function of magnetic field based on the \textit{dV/dI}. The dashed line is a guide to the eye. All \textit{dV/dI} curves are shifted vertically for clarity.}
  \label{Fig-IV-B-GeTe-Nb}
\end{figure*}

The Nb/GeTe-NW/Nb junctions were measured in a magnetic field in two orientations, i.e. parallel to the nanowire axis and perpendicular to the substrate plane. For the perpendicular configuration the critical field $B_{\mathrm{c},\perp}$ was determined to be around 4\,T while for the parallel case we got $B_{\mathrm{c},||} \approx 5$\,T. Figures~\ref{Fig-IV-B-GeTe-Nb}a and \ref{Fig-IV-B-GeTe-Nb}b depict the evolution of \textit{I-V} and corresponding \textit{dV/dI} characteristics of the Nb/GeTe-NW/Nb junction in a perpendicular magnetic field (B$_\perp$) at 0.5\,K. With increasing magnetic field, the supercurrent is suppressed and then disappears, as can be seen in more detail in Figure~\ref{Fig-IV-B-GeTe-Nb}a (inset). The features at around $\pm$1.6 and $\pm$4\,$\mu$A found in the \textit{dV/dI} trace at zero field shift towards zero bias upon increasing the magnetic field. The shift can be attributed to the reduced superconducting gap when a magnetic field is applied. As shown in Figure~\ref{Fig-IV-B-GeTe-Nb}c, we find a monotonous decrease of the critical current I$_{c,\perp}$ with increasing $B_\perp$. A complete suppression of I$_{c,\perp}$ occurs at around 4\,T. A similar monotonous decrease of critical current with $B_\perp$ was observed in planar Nb/Au/Nb\cite{LAngersPRB2008} as well as in semiconductor nanowire based Josephson junctions.\cite{RFrielinghausAPL2010,HYGuenelJAP2012} It can be explained within the framework of a theoretical model for the proximity effect in diffusive narrow-width Josephson junctions.\cite{JCCuevasPRL2007} Within that model the decay of $I_\mathrm{c,\perp}$ can be described by:\cite{JCCuevasPRL2007,LAngersPRB2008} $I_\mathrm{c,\perp}(B)=I_\mathrm{c,\perp}(0)\cdot a e^{-b(BS/\Phi_0)^2}$. Here, $I_\mathrm{C,\perp}(0)$ is the critical current at $B\,=\,0$\,T. The \textit{a} and \textit{b} are fit parameters and \textit{S} is the effective area (80\,nm$\times$90\,nm) of the Nb/GeTe/Nb junction perpendicular to magnetic field. The fit curve from theory (solid line) agrees well with the experimental data for fit parameters \textit{a} and \textit{b} of 1.04 and 0.27, respectively. 

In the case that the applied magnetic field is parallel to the GeTe wire axis (B$_{||}$), the supercurrent is modulated when the magnetic field is increased, as can be seen in Figures~\ref{Fig-IV-B-GeTe-Nb}d and \ref{Fig-IV-B-GeTe-Nb}e. Furthermore, similar to the previous case where $B_\perp$ was applied the peak features in the $dV/dI$ traces shift towards zero bias upon increasing of $B_{||}$. In Figure~\ref{Fig-IV-B-GeTe-Nb}e the ranges where a Josephson supercurrent is present is indicated. The corresponding values of $I_\mathrm{c,||}$ as a function of B$_\mathrm{||}$ are plotted in Figure~\ref{Fig-IV-B-GeTe-Nb}f. One clearly finds an oscillatory behaviour of $I_\mathrm{c,||}$. The oscillation period of $\Delta B_{||}$ is very similar to the period of the sample with normal contacts as discussed before. Since the cross section of the nanowire is very similar to the previous one, we can deduce that oscillations of $I_\mathrm{c,||}$ are periodic with a single magnetic flux quantum $\Phi_0=h/e$. We interpret the behaviour by a Josephson supercurrent which is carried by coherent closed-loop states in the surface accumulation layer at the GeTe nanowire surface. As illustrated in Fig.~\ref{Fig-SEM-GeTe}e, bottom, the supercurrent originates from phase-coherent Andreev retro-reflections of tubular hole and electron states at both GeTe/Nb interfaces.\cite{IOKulik1969} In each cycle a Cooper pair is effectively transported from one Nb electrode to the other one. In the measurements of GeTe nanowire with normal contacts we found that the resistance was modulated with a magnetic flux quantum because of the periodic change of the number of occupied transport channels. In case of superconducting electrodes, the magnitude of the Josephson supercurrent is also considered to be determined by the number of hole and electron transport channels, hence a flux periodic modulation is expected as well. As a matter of fact, a magnetic flux modulation was found before in junctions based on GaAs/InAs core/shell nanowires.\cite{OGuelPRB2014,FHaasSST2018} However, in that case no clear supercurrent was observed and a period of $\Phi_0/2$ was found.   

\section{Conclusion}

In summary, magnetotransport of GeTe nanowire-based devices have been investigated in a four-terminal configuration. The temperature dependent resistance of the normal metal contacted Au/GeTe-nanowire/Au devices reveals that GeTe exhibits a semiconducting behavior until 0.4\,K. At low temperatures universal conductance fluctuations were observed which allowed us to extract the phase-coherence length $l_\phi$ being as large as 200\,nm at 0.5\,K. The weak antilocalization feature observed around zero magnetic field indicates the presence of spin-orbit coupling in the valence band. Even more, when a parallel magnetic field is applied regular Aharonov--Bohm type oscillations are found, which are attributed to the formation of a hole accumulation layer at the nanowire surface. In the case of Nb/GeTe-NW/Nb junctions, we observed a critical supercurrent of about $0.2$\,$\mu$A at 0.4\,K. The temperature dependence of $I_\mathrm{c,NW}$ can be explained in the framework of a diffusive junction. With increasing a perpendicular magnetic field the critical current decreases monotonously. This feature can be explained within models covering the small junction limit. For magnetic fields applied parallel to the nanowire axis regular oscillations of the critical current are observed which are attributed to a supercurrent carried by the surface accumulation layer in the GeTe nanowires. 

Our investigations showed that distinct phase-coherent phenomena can be observed in GeTe nanowire structures. The presence of spin-orbit coupling in combination with the superconducting proximity effect make these nanowires very attractive for applications in the field of quantum computation. Furthermore, it would be very interesting to measure the bulk Rashba effect in connection with the ferroelectric properties under a gate bias.    

\section{Experimental Details}

\textit{GeTe nanowire synthesis}: The synthesis of GeTe nanowires are achieved by a Au-catalyzed vapor-liquid-solid growth. Firstly, the Si(100) substrates with a native oxide were cleaned with acetone, iso-propyl alcohol, and deionized water in an ultrasonic cleaning bath, and then treated in a piranha solution to remove organic residues. After that, the substrates were immersed in Au nanoparticles solution for a few min and rinsed with deionized water. Finally, bulk GeTe (99.99\%, Sigma-Aldrich) was evaporated at the center of a horizontal tube furnace, and the reaction product was collected downstream on a Si/SiO$_2$ substrate covered with colloidal Au nanoparticles. Specifically, the furnace was evacuated and purged three times with Ar gas. Then it was heated to 400\,$^\circ$C and persisted for 8 h. During this program, the Ar flow rate and the pressure in the quartz tube were about 140\,sccm and\,10 Torr, respectively.

\textit{GeTe-based device fabrication}: In order to contact a single GeTe nanowire with Nb or Au electrodes, the as-grown GeTe NWs were transferred to a Si/SiO$_2$/Si$_3$N$_4$ substrate with predefined markers for electron beam lithography. A two-layer (copolymer/950K) polymethyl methacrylate (PMMA) resist system was adopted to realized an ideal shape of Nb or Au leads. In order to obtain a transparent interface between electrodes and GeTe, the samples were exposed to an oxygen plasma to remove resist residues on the contact area. Before the deposition of the electrodes on a single nanowire in a magnetron sputter chamber with a DC power of 250 W, the samples were lightly cleaned by 45\,s Ar$^+$ plasma milling to remove native oxidation layers on the GeTe surface.

\textit{Morphology and composition characterizations}: The morphology and chemical composition of the GeTe nanowires was determined by SEM and EDX, respectively. In addition, a cross sectional specimen has been prepared by focused ion beam techniques. High-angle annular dark field imaging in the STEM has been employed to investigate the cross sectional shape of the nanowires. In addition, the crystalline structure has been investigated by x-ray diffraction and Raman scattering. 

\textit{Magnetotransport measurements}: The temperature and magnetic field dependent \textit{I-V} characteristics and \textit{dV/dI} curves were measured using a standard lock-in technology in a $^3$He cryostat with a base temperature of 0.4\,K and a magnetic field range from $-8$ to 8\,T. A four-terminal current driven geometry is employed to directly measuring the voltage drop across the inner section of GeTe nanowire between the nearest internal electrodes. The external pair of electrodes was used for current bias.\cite{JZZhangAnnalPhys2020} The differential resistance (\textit{dV/dI}) is gained by superimposing a small AC current of 10\,nA at 9.4\,Hz on a DC bias current. Note that the applied bias current should be as small as possible to avoid electron heating and damage of the GeTe-based nanodevices.

\section*{Acknowledgements} 
We thank Peter Sch\"{u}ffelgen, Tobias Schmitt, and Michael Schleenvoigt for helpful discussions, Herbert Kertz for technical assistance and Stefan Trellenkamp and Lidia Kibkalo for assistance during prepration and characterization. Furthermore, we thank the Helmholtz Nano Facility (HNF) for help with devices fabrication in its clean room facilities.\cite{HNFJournalLSF2017} This work was financially supported by the International Postdoctoral Exchange Fellowship Program (No.20161008), the German Deutsche Forschungsgemeinschaft (DFG) under the priority program SPP1666 ``Topological Insulators'', the Helmholtz Association the ``Virtual Institute for Topological Insulators'' (VITI, VH-VI-511), by the Deutsche Forschungsgemeinschaft (DFG, German Research Foundation) under Germany’s Excellence Strategy - Cluster of Excellence Matter and Light for Quantum Computing (ML4Q) EXC 2004/1 - 390 534 769, the National Natural Science Foundation of China (Grant No. 61504156), the National Key R\&D Program of China (Grants No. 2017YFA0303403 and No. 2018YFB0406500), and the  Projects of Science and Technology Commission of Shanghai Municipality (Grant No. 18JC1412400, \\

\section*{Author contributions}
J.Z. J.K., and D.R. performed the transport experiments, J.Z., A.R.J, J.K., D.R., and G.P. prepared the samples in the cleanroom, P.L.T and J.G.L. fabricated the nanowires by VLS growth, M.L., A.R.J, J.Z., and J.G.L. performed the structural analysis. Y.Z. A.R.J., and T.S. wrote the paper. All authors contributed to the discussions. 

\section*{Competing interests:} The authors declare no competing interests.
\\

\section*{Keywords}

GeTe nanowires, Josephson junction, four-terminal measurement, proximity effect, quantum transports

\section*{Additional Information}

\textbf{Supplementary Information} accompanies this paper at http://www.nature.com/naturecommunications 
\\

\textbf{Corresponding author:}
Correspondence to T.S.. \\ 

\section*{Data availability}
The data that support the findings of this study is available from
the authors on a reasonable request.

\end{document}